\documentclass[
   pre,
   superscriptaddress,
   twocolumn,
   eqsecnum,
   preprintnumbers,
   byrevtex]{revtex4}
\usepackage{amsfonts}
\usepackage{amssymb}
\usepackage{amsmath}
\usepackage{color}
\usepackage{graphicx}
\usepackage{subfigure}
\usepackage{time}
\usepackage[pdftex,colorlinks=true]{hyperref}

\newcommand{\kk}{\mathbf{k}}
\newcommand{\qq}{\mathbf{q}}

\usepackage{lipsum}

\newcounter{mycounter}

\begin{document}
%
%%%%%%%%%%%%%%%%%%%%%%%%%%%%%%%%%%%%%%%%%%%%%%%%%%%%%%%%%%%%%%%%%%%%%%
%
% titlepage
%
\preprint{LA-UR-11-02247} %%%; \today\ \now}
\title{Lindhard function of a $d$-dimensional Fermi gas}

\author{Bogdan Mihaila}
\email{bmihaila@lanl.gov}
\affiliation{%Materials Science and Technology Division,
   Los Alamos National Laboratory,
   Los Alamos, NM 87545}

\begin{abstract}
     We review in detail the derivation of the dielectric response function of a noninteracting system of spin-$1/2$ fermions in the random-phase approximation. 
     Results for the response function of a Fermi gas in one, two and three dimensions can be obtained in closed form, and represent the baseline for 
     developing a pedagogical 
     understanding of the effect of correlations on the response functions in interacting systems of fermions.
\end{abstract}

\maketitle

%
%%%%%%%%%%%%%%%%%%%%%%%%%%%%%%%%%%%%%%%%%%%%%%%%%%%%%%%%%%%%%%%%%%%%%%
%

\section{Introduction}
\label{sec:intro}

The study of quantum systems with low spatial dimensions represents an active field of study that is being pursued aggressively in the context of  emergent phenomena and has important implications for both fundamental and applied physics. For instance, the reduced dimensionality of the system is relevant to the properties of thin films and surfaces~\cite{stern}, and for the understanding of high-temperature cuprate superconductors~\cite{2d_peter}. This in turn poses classroom challenges, aiming at exposing students to the array of formal techniques required to study such systems, with a special emphasize on linear response theory aimed at studying the response of the quantum system to a perturbation~\cite{teaching_smith}.

In quantum many-body systems
the role played by correlations and quantum fluctuations is greatly affected by the reduced dimensionality of the system. 
%At low density, such as in quantum gases, the details of the inter-particle interactions are greatly suppressed and the system properties have an universal character. 
%In this context, we can build on the studies of scattering in quantum mechanics as a function of the number of spatial dimensions~\cite{1d_scattering78,1d_scattering08,2d_scattering80,2d_scattering86}. 
Understanding the role played by collective excitations  in calculating the response to a perturbation of a many-body system represents a fundamental problem in many-body theory~\cite{pines,fetter}. The special case of a homogeneous gas of interacting spin-$1/2$ fermions has important applications ranging from condensed matter and ultracold atom gases to nuclear and high-energy physics. Furthermore, the response of an interacting system can be represented as the response of a noninteracting system of particles to an effective self-consistent field. This is true in particular for weakly-interacting systems of particles. Therefore, a classic problem in many-body theory is the calculation of the response function for a system of noninteracting spin-$1/2$ fermions (e.g. electrons). 

The intent of this paper is pedagogical. 
The response function of a Fermi gas in three, two and one dimensions can be obtained in closed form, and this derivation represents a useful classroom exercise. In addition,  
the Fermi gas results are intended as the baseline for studying the dielectric response function in a $d$-dimensional Penn-model semiconductor~\cite{penn}, perhaps the simplest semi-realistic demonstration of changes introduced by the presence of a gap in the single-particle energy dispersion relation~\cite{penn_paper}. The latter will be discussed elsewhere.

%
%%%%%%%%%%%%%%%%%%%%%%%%%%%%%%%%%%%%%%%%%%%%%%%%%%%%%%%%%%%%%%%%%%%%%%
%

\section{Dielectric function in RPA}
\label{sec:basics}

Consider the interaction of an electrostatic field, $\Phi^\mathrm{ext}(\mathbf{r},\omega)$, with a system of fermions. Then, the external potential will induce a charge distribution, $\rho^\mathrm{ind}(\mathbf{r},\omega)$, which in turn produces a potential $\Phi^\mathrm{ind}(\mathbf{r},\omega)$. The total electrostatic potential $\Phi^\mathrm{tot}(\mathbf{r},\omega)$ is given by the sum of the external potential and that associated with the induced charge distribution, i.e.
\begin{align}
   \Phi^\mathrm{tot}(\mathbf{r},\omega)
   = & \
   \Phi^\mathrm{ext}(\mathbf{r},\omega)
   +
   \Phi^\mathrm{ind}(\mathbf{r},\omega)
   \>.
\end{align}
Similarly, the total electric charge, $\rho^\mathrm{tot}(\mathbf{r},\omega)$, is given by the superposition of the charge distribution of the noninteracting target, $\rho^\mathrm(\mathbf{r},\omega)$, and the induced charge distribution, i.e.
\begin{align}
   \rho^\mathrm{tot}(\mathbf{r},\omega)
   = & \
   \rho(\mathbf{r},\omega)
   +
   \rho^\mathrm{ind}(\mathbf{r},\omega)
   \>.
\end{align}
The individual external, induced and total charge distributions and electrostatic potentials obey Poisson equations of the form
\begin{align}
   - \, \nabla^2 \Phi = & \ 4\pi \, \rho
   \>.
\end{align}
In a linear-response approximation, we have
\begin{align}
   \rho^\mathrm{ind}(\mathbf{r},\omega)
   =
   e^2
   \!\! \int
   \mathrm{d}^d r' \
   \chi(\mathbf{r},\mathbf{r}',\omega) \
   \Phi^\mathrm{tot}(\mathbf{r}',\omega) \
   \>,
\end{align}
where the electron response function is calculated in the random-phase approximation (RPA), as
\begin{align}
   \chi(\mathbf{r},\mathbf{r}',\omega)
   =
   \lim_{\eta \rightarrow 0^+}
   \sum_{ij} &
   \frac{f_i - f_j}{\hbar ( \omega - \omega_{ij} ) + \mathrm{i} \eta} \,
   \\ \notag & \quad \times \
   \psi_i^*(\mathbf{r}) \psi_j^*(\mathbf{r}') \psi_j(\mathbf{r})\psi_i(\mathbf{r}')
   \>,
\end{align}
or, in momentum space,
\begin{align}
   \chi(\mathbf{q},\mathbf{q}';\omega)
   =
   \lim_{\eta \rightarrow 0^+}
   \sum_{ij} &
   \frac{f_i - f_j}{\hbar ( \omega - \omega_{ij} ) - \mathrm{i} \eta} \
\label{eq:phi}
   \\ \notag & \quad \times \
   \langle i | e^{\mathrm{i}\mathbf{q} \cdot \mathbf{r}'} | j \rangle
   \langle j | e^{-\mathrm{i}\mathbf{q}' \cdot \mathbf{r}''} | i \rangle
   \>.
\end{align}
Here, $\varepsilon_i$ and $\psi_i(\mathbf{r})$ are single-particle energies and wave functions, and $f_i$ are occupation functions of the single-particle states.
We have also introduced the notation, $\hbar \, \omega_{ij} = \varepsilon_i - \varepsilon_j$, and  the matrix elements are calculated as
\begin{align}
   \langle i | e^{\mathrm{i}\mathbf{q} \cdot \mathbf{r}} | j \rangle
\label{eq:matel}
   =
   \int \mathrm{d}^d r \
   \psi_i^*(\mathbf{r}) \
   e^{\mathrm{i}\mathbf{q} \cdot \mathbf{r}} \
   \psi_j(\mathbf{r})
   \>.
\end{align}
For uniform systems, we have
\begin{align}
   &
   \Phi^\mathrm{ind}(\mathbf{q},\omega)
   =
   e^2 \,
   \mathcal{V}_d(q)
   \int \frac{\mathrm{d}^d q'}{(2\pi)^d} \
   \chi(\mathbf{q},\mathbf{q}';\omega) \
   \Phi^\mathrm{tot}(\mathbf{q}',\omega)
   \>,
\end{align}
and $\mathcal{V}_d(q)$ is the Fourier transform of the coulomb potential in \emph{d}~dimensions, i.e.
\begin{equation}
   \mathcal{V}_d(q) =
   \!\! \int \mathrm{d}^d {\rm x} \ \frac{e^{\mathrm{i} \mathbf{q} \cdot \mathbf{x}}}{\rm x}
   =
   \left \{
   \begin{array}{ll}
      - 2 \gamma_E + \ln(1/q^2) \>, & \mathrm{if}\ d = 1 \>,
      \\
      2\pi / q \>, & \mathrm{if}\ d = 2 \>,
      \\
      4\pi / q^2 \>, & \mathrm{if}\ d = 3 \>,
   \end{array}
   \right .
\end{equation}
with the Euler constant, $\gamma_E = 0.577216 \cdots$.
Therefore, the dielectric function in RPA reads
\begin{equation}
   \varepsilon(\mathbf{q},\mathbf{q}';\omega)
   =
   \delta_{\mathbf{q},\mathbf{q}'} + e^2 \, \mathcal{V}_d(q) \, \chi(\mathbf{q},\mathbf{q}';\omega)
   \>.
\end{equation}

%
%%%%%%%%%%%%%%%%%%%%%%%%%%%%%%%%%%%%%%%%%%%%%%%%%%%%%%%%%%%%%%%%%%%%%%
%

\section{Noninteracting (Fermi gas) model}
\label{sec:fermi}

A system of noninteracting spin-$1/2$ fermions is commonly referred to as a Fermi Gas (FG) system. In this system, one-particle states are plane waves characterized by momentum $\kk$, spin $\sigma = \pm 1$, corresponding to the usual spin-up ($\uparrow$) and spin-down ($\downarrow$) fermions, and one-particle energies, $\epsilon_{\kk} = \hbar^2 k^2 / (2m)$. All states below momentum Fermi, $k_F$, are occupied. Hence, the occupation of the one-particle states is given by
\begin{equation}
n_{\kk} = \Theta(k_F - k)
\>.
\end{equation}
In the following the sum over states $k$ is understood as an integral over all one-particle states, i.e
\begin{equation}
\sum_k \ \rightarrow \ \sum_\sigma \int \frac{\mathrm{d}^dk}{(2\pi)^d}
\>.
\end{equation}
The Fermi gas density in $d$~dimensions is calculated as follows
\begin{align}
N_1 & = 2 \int_{k \le k_F} \frac{\mathrm{d}k}{2\pi}
= \frac{2 k_F}{\pi}
\>,
\\
N_2 & = 2 \int_{k \le k_F} \frac{\mathrm{d}^2k}{(2\pi)^2}
= \frac{k_F^2}{2\pi}
\>,
\\
N_3 & = 2 \int_{k \le k_F} \frac{\mathrm{d}^3k}{(2\pi)^3}
= \frac{k_F^3}{3\pi^2}
\>.
\end{align}
As a matter of convenience we express momenta in terms of the Fermi momentum, $k_F$, and energies in terms of the Fermi energy, $\epsilon_F = \hbar^2 k_F^2 / ( 2m)$.

Assuming Fermi-gas-like occupation functions, 
$
   f_\mathbf{k}
   = n_\mathbf{k} 
$, 
%\begin{align}
%   = \Theta(k_F - k)
%   = \left \{
%   \begin{array}{ll}
%      1 \>, & \mathrm{if}\ k < k_F\>,
%      \\
%      0 \>, & \mathrm{otherwise} \>,
%   \end{array}
%   \right .
%\end{align}
the response function reads
\begin{align}
   \chi(\mathbf{q},\mathbf{q}';\omega)
   = &
   \ 2 \,
   \int \frac{\mathrm{d}^d k_i}{(2\pi)^d} \ n_{\mathbf{k}_i}
   \\ \notag & \times
   \int \!\! \mathrm{d}^d k_j \ (1 - n_{\mathbf{k}_j})
   \lim_{\eta \rightarrow 0^+}
   \frac{\mathcal{F}(\mathbf{k}_i,\mathbf{k}_j;\mathbf{q},\mathbf{q}')}
        {\hbar ( \omega - \omega_{ij} )
         + \mathrm{i} \eta}
   \>,
\end{align}
where $\theta(x)$ is the Heaviside function and $\mathcal{F}(\mathbf{k}_i, \mathbf{k}_j; \mathbf{q})$ depends on the ground-state model. 
The factor 2 is the degeneracy factor corresponding to a system of spin-$1/2$ fermions.

The matrix elements~\eqref{eq:matel} corresponding to the case of a Fermi gas system are given by 
\begin{align}
   \frac{1}{(2\pi)^d } \,
   \langle j | e^{\mathrm{i}\mathbf{q} \cdot \mathbf{r}} | i \rangle
 \label{eq:matel_fg}
   =
   \delta^d (\mathbf{q}+\mathbf{k}_i-\mathbf{k}_j)
   \>.
\end{align}
Therefore, we find that
\begin{align}
   \mathcal{F}(\mathbf{k}_i, & \mathbf{k}_j; \mathbf{q},\mathbf{q}')
\label{eq:F_fg}
   = \
   \delta^d_{\mathbf{q}',\mathbf{q}} \,
   \Bigl (
   \delta^d_{\mathbf{k}_j,\mathbf{k}_i+\mathbf{q}}
   \, - \,
   \delta^d_{\mathbf{k}_j,\mathbf{k}_i-\mathbf{q}}
   \Bigr )
   \>,
\end{align}
and the response (Lindhard) function for the case of the Fermi gas is local. We obtain
\begin{align}\label{response}
   \chi_0(\mathbf{q},\omega)
   = & \,
   \frac{2}{\hbar}
   \int \!\! \frac{\mathrm{d}^d k}{(2\pi)^d} \ n_{\mathbf{k}} \
      (1 - n_{\mathbf{k}+\mathbf{q}})
   \\ \notag & \quad \times
   \Bigl [
   \frac{1}
        {\omega - \omega_{\mathbf{kq}}
         + \mathrm{i} \eta}
   -
   \frac{1}
        {\omega + \omega_{\mathbf{kq}}
         - \mathrm{i} \eta}
   \Bigr ]
   \>,
\end{align}
where we have introduced the notation
\begin{align}
   \hbar \, \omega_{\mathbf{kq}} = & \
   \varepsilon_{|\mathbf{k}+\mathbf{q}|} - \varepsilon_{k}
   = \frac{\hbar^2}{2m} \Bigl ( q^2 + 2 \, \mathbf{q} \cdot \mathbf{k} \Bigr )
   \>.
\end{align}

%
%%%%%%%%%%%%%%%%%%%%%%%%%%%%%%%%%%%%%%%%%%%%%%%%%%%%%%%%%%%%%%%%%%%%%%
%

\section{Static response functions}
\label{sec:static}

The static case corresponds to the zero-energy transfer limit, $\omega=0$. This case was discussed in detail by Kittel~\cite{kittel} and we will follow his approach here.

For $\omega=0$ in Eq.~\eqref{response}, it is conventional to introduce the notation $F(\qq) = - \, \chi(\mathbf{q},\omega=0)$.
%In the RPA, only unoccupied one-particle states are allowed as intermediate states in the perturbation calculation. 
We have:
\begin{align}
F(\qq) 
& = 
   2
   \!\! \int_{k \le k_F} \frac{\mathrm{d}^d k}{(2\pi)^d} \ 
\Bigl [ \frac{n_{\kk} (1- n_{\kk+\qq})}{\epsilon_{\kk+\qq} - \epsilon_{\kk}}
- \frac{(1- n_{\kk-\qq})n_{\kk} }{\epsilon_{\kk} - \epsilon_{\kk-\qq} } \Bigr ]
\notag \\
& = 
   2
  \!\! \int_{k \le k_F} \frac{\mathrm{d}^d k}{(2\pi)^d} \ 
\Bigl [ \frac{n_{\kk} (1- n_{\kk+\qq})}{\epsilon_{\kk+\qq} - \epsilon_{\kk}}
- \frac{(1- n_{\kk})n_{\kk+\qq} }{\epsilon_{\kk+\qq} - \epsilon_{\kk}} \Bigr ]
\notag \\
& = 
   2
   \!\! \int_{k \le k_F}  \frac{\mathrm{d}^d k}{(2\pi)^d} \ 
\frac{n_{\kk} - n_{\kk+\qq}}{\epsilon_{\kk+\qq} - \epsilon_{\kk}}
\>,
\end{align}
where we changed variables in the second term based on the fact that the integral is over all states $k$.
By the same argument, we can write
\begin{equation}
F(\qq) 
= 
   2
   \int_{k \le k_F}  \ \Bigl ( \frac{n_{\kk}}{\epsilon_{\kk+\qq} - \epsilon_{\kk}}
 - \frac{n_{\kk}}{\epsilon_{\kk} - \epsilon_{\kk-\qq}} \Bigr )
\>,
\end{equation}
or
\begin{equation}
\frac{\hbar^2}{2m} \, F(\qq) 
= 
2 \!\! \int_{k \le k_F} \frac{\mathrm{d}^dk}{(2\pi)^d} \
\Bigl ( \frac{1}{q^2 + 2 \, \kk \cdot \qq}
 + \frac{1}{q^2 - 2 \, \kk \cdot \qq} \Bigr )
\>.
\end{equation}
Using rescaled coordinates, i.e. by letting $k \rightarrow k/k_F$, we obtain
\begin{equation}
F(\qq) 
= 
\frac{2 k_F^d}{\epsilon_F} \int_{k \le 1} \frac{\mathrm{d}^dk}{(2\pi)^d} \
\frac{2 q^2}{q^4 - 4 (\kk \cdot \qq)^2}
\>.
\end{equation}
We will discuss now the three-dimensional realizations of the static response function corresponding to $d=1$, 2 and 3. We have:

%
%%%%%%%%%%%%%%%%%%%%%%%%%%%%%%%%%%%%%%%%%%%%%%%%%%%%%%%%%%%%%%1
\begin{figure}[t]
   \centering
   \includegraphics[width=0.9\columnwidth]{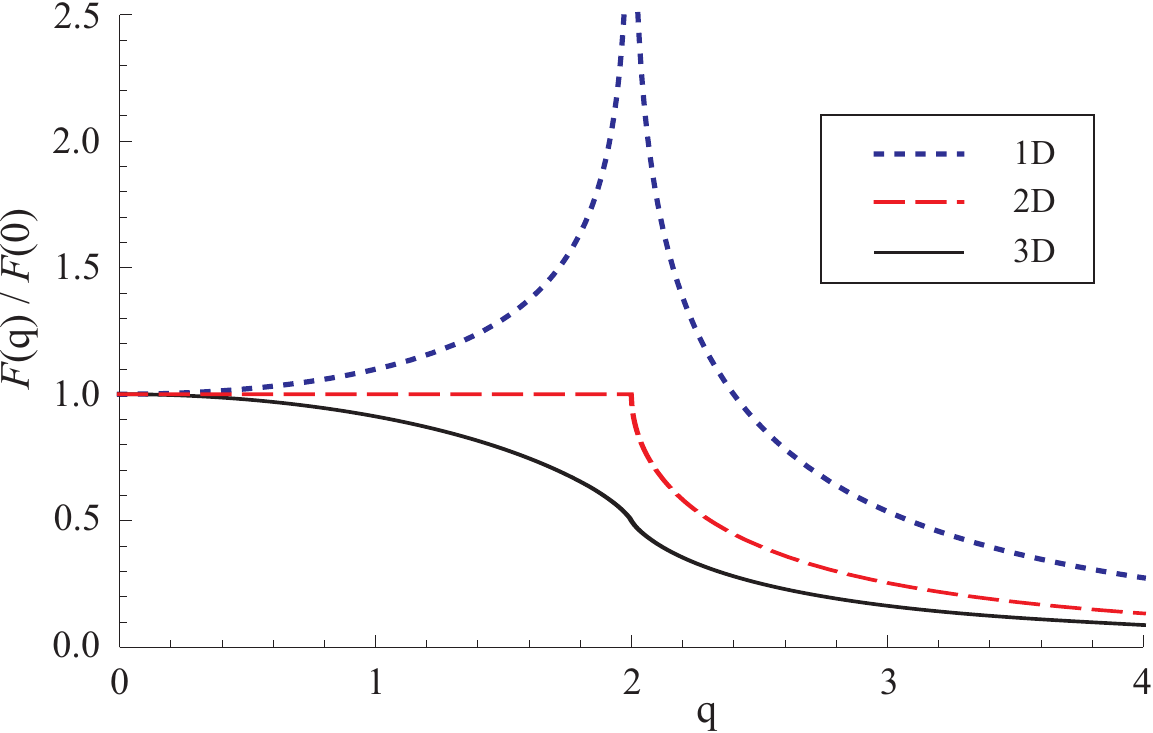}
   \caption{\label{Fig_stat} (Color online) Momentum dependence of the $d$-dimensional static response functions, $F(q) = - \, \chi(q,\omega=0)$, corresponding to zero energy transfer, $\omega=0$. Here, momenta are expressed in rescaled units, i.e. in units of the Fermi momentum, $k_F$.}
\end{figure}
%
%%%%%%%%%%%%%%%%%%%%%%%%%%%%%%%%%%%%%%%%%%%%%%%%%%%%%%%%%%%%%

%
\begin{list}{\labelitemi}{\leftmargin=1em}
%%%%%%%%%%%%%%%%%%%%%%%%%%%%%%%%%%%%%%%%%%%%%%%%%%%%%%%%%%%%%%%%%%%%%%
\item $d=3$ case:
We have
\begin{align*}
F_3(q) 
= &
\frac{k_F^3}{\pi^2 \epsilon_F}  
\int_0^1 k^2 \, \mathrm{d}k \int_{-1}^1 
\frac{\mathrm{d}\mu}{q^2 - 4 k^2 \mu^2}
\\
= &
\frac{N_3}{\epsilon_F} \,
\frac{3}{2q}  
\int_0^1 k \, \mathrm{d}k \,
\log \Bigl | \frac{q+2k}{q-2k} \Bigr |
\>,
\end{align*}
which gives
\begin{equation}
F_3(q) 
= 
\frac{N_3}{\epsilon_F} \, \frac{3}{4} \,
\Bigl [ 
1
+ \frac{1 - (\frac{1}{2}q)^2}{q} \, \log \Bigl | \frac{1 + \frac{1}{2}q}{1 - \frac{1}{2}q} \Bigr |
\Bigr ]
\>.
\end{equation}

%
%%%%%%%%%%%%%%%%%%%%%%%%%%%%%%%%%%%%%%%%%%%%%%%%%%%%%%%%%%%%%%%%%%%%%%
\item $d=2$ case:
We have
\begin{equation}
F_2(q) 
= 
\frac{k_F^2}{\pi^2 \epsilon_F}  \int_0^1 k \, \mathrm{d}k \int_0^{2\pi} 
\frac{\mathrm{d}\phi}{q^2 - 4 k^2 \cos^2 \phi}
\>.
\end{equation}
For $q \ge 2$, the above gives
\begin{align}
F_2(q>2) 
= & \,
\frac{N_2}{\epsilon_F} \, \frac{4}{q}
\int_0^1  \frac{k \, \mathrm{d}k}{\sqrt{q^2 - 4k^2}}
\notag \\
= & \,
\frac{N_2}{\epsilon_F} \
\Bigl ( 1- \sqrt{1 - (2/q)^2}
\Bigl ) 
\>,
\end{align}
whereas for $q<2$, we obtain
\begin{equation}
F_2(q<2) 
= \frac{N_2}{\epsilon_F} \, \frac{4}{q}  \int_0^{\frac{1}{2}q}  \frac{k \, \mathrm{d}k}{\sqrt{q^2 - 4k^2}}
= \frac{N_2}{\epsilon_F} 
\>,
\end{equation}
Hence, we obtain:
\begin{align}
F_2(q) 
= 
\frac{N_2}{\epsilon_F} \,  
\Bigl [ 
1 \, - \, 
\Theta ( q - 2 ) \ \sqrt{1 - (2/q)^2} \
\Bigr ]
\>.
\end{align}

%
%%%%%%%%%%%%%%%%%%%%%%%%%%%%%%%%%%%%%%%%%%%%%%%%%%%%%%%%%%%%%%%%%%%%%%
\item $d=1$ case:
We have
\begin{align}
F_1(q) 
= &
\frac{2 k_F}{\pi \epsilon_F}  \int_{-1}^1 \frac{\mathrm{d}k}{q^2 - 4 k^2}
\>,
\end{align}
which gives
\begin{equation}
F_1(q) 
= 
\frac{N_1}{\epsilon_F} \, \frac{1}{2q} \,  
\log \Bigl | \frac{1 + \frac{1}{2}q}{1 - \frac{1}{2}q} \Bigr |
\>.
\end{equation}

\end{list}
In Fig.~\ref{Fig_stat}, we depict the momentum dependence of the above static response functions, $F_1(q)$, $F_2(q)$, and  $F_3(q)$.

%
%%%%%%%%%%%%%%%%%%%%%%%%%%%%%%%%%%%%%%%%%%%%%%%%%%%%%%%%%%%%%%%%%%%%%%
%

\section{Lindhard functions}
\label{sec:Lindhard}

To calculate the Lindhard function, i.e.
\begin{align}
\chi(\qq,\omega)
= \frac{2}{\hbar} \, \int_{k \le k_F} & \frac{\mathrm{d}^dk}{(2\pi)^d} \
n_\kk (1 - n_{\kk+\qq}) \, 
\\ & \notag \times \,
\Bigl ( \frac{1}{\omega - \omega_{\kk \qq} + \mathrm{i} \eta}
- \frac{1}{\omega + \omega_{\kk \qq} - \mathrm{i} \eta} \Bigr )
\>,
\end{align}
we use the formal identity
\begin{equation}
\frac{1}{\omega \pm \mathrm{i}\eta}
= \mathcal{P} \, \frac{1}{\omega} \ \mp \ \mathrm{i} \pi \delta(\omega)
\>.
\end{equation}
Therefore, we have 
\begin{equation}
\chi(\qq,\omega) = \mathcal{R}e \, \chi(\qq,\omega) \, + \, \mathrm{i} \, \mathcal{I}m \, \chi(\qq,\omega)
\>,
\end{equation}
with
\begin{equation}
\mathcal{R}e \, \chi(\qq,\omega)
\label{eq:re_0}
= \frac{2}{\hbar} \ \mathcal{P} \! \int \frac{\mathrm{d}^dk}{(2\pi)^d} \,
n_\kk (1 - n_{\kk+\qq}) \, \frac{2\omega_{\kk \qq}}{\omega^2 - \omega_{\kk \qq}^2}
\>,
\end{equation}
and
\begin{align}
\mathcal{I}m \, \chi(\qq,\omega)
\label{eq:im_0}
= - \frac{2\pi}{\hbar} \int & \frac{\mathrm{d}^dk}{(2\pi)^d} \
n_\kk (1 - n_{\kk+\qq}) \, &
\\ \notag & 
\times \Bigl [ \delta(\omega - \omega_{\kk \qq}) + \delta(\omega + \omega_{\kk \qq})
\Bigr ]
\>,
\end{align}
where we introduce the notation
\begin{equation}
\hbar \omega_{\kk \qq}
=
\epsilon_{\kk+\qq} - \epsilon_{\kk}
= \frac{\hbar^2}{2m} \Bigl ( q^2 + 2 \, \kk \cdot \qq \Bigr )
\>.
\end{equation}

%
%%%%%%%%%%%%%%%%%%%%%%%%%%%%%%%%%%%%%%%%%%%%%%%%%%%%%%%%%%%%%%%%%%%%%%
%

\subsection{Real part of the Lindhard function}

In Eq.~\eqref{eq:re_0}, we observe that the product $n_\kk (1 - n_{\kk+\qq})$ is symmetric under the interchange $\kk \leftrightarrow \kk + \qq$, whereas $\omega_{\kk \qq}$ is odd. Hence the second term in Eq.~\eqref{eq:re_0} vanishes and we obtain
\begin{equation}
\mathcal{R}e \, \chi(\qq,\omega)
= \frac{2}{\hbar} \, \mathcal{P} \! \int_{k \le k_F} \frac{\mathrm{d}^dk}{(2\pi)^d} \,
\frac{2\omega_{\kk \qq}}{\omega^2 - \omega_{\kk \qq}^2}
\>.
\end{equation}
Using rescaled coordinates and after introducing the notation, $\nu=\hbar \omega / \epsilon_F$, we obtain
\begin{align}
& \mathcal{R}e \, \chi(\qq,\nu)
\\ \notag & 
= \frac{2 k_F^d}{\epsilon_F} \ 
\mathcal{P} \! \int_{k \le 1} \frac{\mathrm{d}^dk}{(2\pi)^d} 
\Bigl ( \frac{1}{q_-^2 - 2 \kk \cdot \qq}
- \frac{1}{q_+^2 + 2 \kk \cdot \qq}
\Bigr )
\>,
\end{align}
with $q_{\pm}^2 = \nu \pm q^2$.
We can also write:
\begin{equation}
\mathcal{R}e \, \chi(\qq, \nu) = \mathcal{R}e \, \chi^{(-)}(\qq, \nu) 
-  \mathcal{R}e \, \chi^{(+)}(\qq, \nu) 
\>,
\end{equation}
where the two contributions are
\begin{equation}
\mathcal{R}e \, \chi^{(\pm)}(\qq, \nu) =
\frac{2 k_F^d}{\epsilon_F} \ \mathcal{P} \! \int_{k \le 1} \frac{\mathrm{d}^dk}{(2\pi)^d} \, 
\frac{1}{q_\pm^2 \pm \qq \cdot \kk}
\>.
\end{equation}
We note the relationship:
\begin{equation}
\mathcal{R}e \, \chi(\qq, \nu=0) = - F(\qq)
\>.
\end{equation}
We specialize now to the three-dimensional realizations corresponding to $d=1$, 2 and 3. We have:

\begin{list}{\labelitemi}{\leftmargin=1em}
%%%%%%%%%%%%%%%%%%%%%%%%%%%%%%%%%%%%%%%%%%%%%%%%%%%%%%%%%%%%%%%%%%%%%%
\item $d=3$ case:
We have:
\begin{align}
\mathcal{R}e \, & \chi_3^{(\pm)}(q, \nu)
= \frac{2 k_F^3}{(2\pi)^2\epsilon_F} \ \mathcal{P} \! \int_0^1 k^2 \, \mathrm{d}k \,
\int_{-1}^{1} \frac{\mathrm{d}\mu}{q_\pm^2 \pm 2 q k \mu}
\notag \\ & 
= \frac{N_3}{\epsilon_F} \, \frac{3}{4q} \int_0^1 k \, \mathrm{d}k \,
\log \Bigl | \frac{q_\pm^2 + 2 q k}{q_\pm^2 - 2 q k} \Bigl |
\>,
\end{align}
which gives
\begin{align}
\mathcal{R}e \, & \chi_3^{(\pm)}(q, \nu) 
\\ \notag & 
= \frac{N_3}{\epsilon_F} \, \frac{3}{4} \, 
\Bigl [ 
\frac{q_\pm^2}{2q^2} + \frac{4q^2 - q_\pm^4}{8q^3} \ \log \Bigl | \frac{1 + q_\pm^2/(2q)}{1 - q_\pm^2/(2q)} \Bigr |
\Bigr ]
\>.
\end{align}
Hence, we obtain the well-known 3D Lindhard function~\cite{lindhard}:
\begin{align}
\mathcal{R}e \, \chi_3(q, \nu) 
= & 
\frac{N_3}{\epsilon_F} \, \frac{3}{4} \, 
\biggl [  \,
- 1 
+ \frac{4q^2 - q_-^4}{8q^3} \ \log \Bigl | \frac{1 + q_-^2/(2q)}{1 - q_-^2/(2q)} \Bigr |
\notag \\ & 
- \frac{4q^2 - q_+^4}{8q^3} \ \log \Bigl | \frac{1 + q_+^2/(2q)}{1 - q_+^2/(2q)} \Bigr |
\, \biggr ]
\>.
\end{align}

%
%%%%%%%%%%%%%%%%%%%%%%%%%%%%%%%%%%%%%%%%%%%%%%%%%%%%%%%%%%%%%%1
\begin{figure}[t]
   \centering
   \includegraphics[width=0.9\columnwidth]{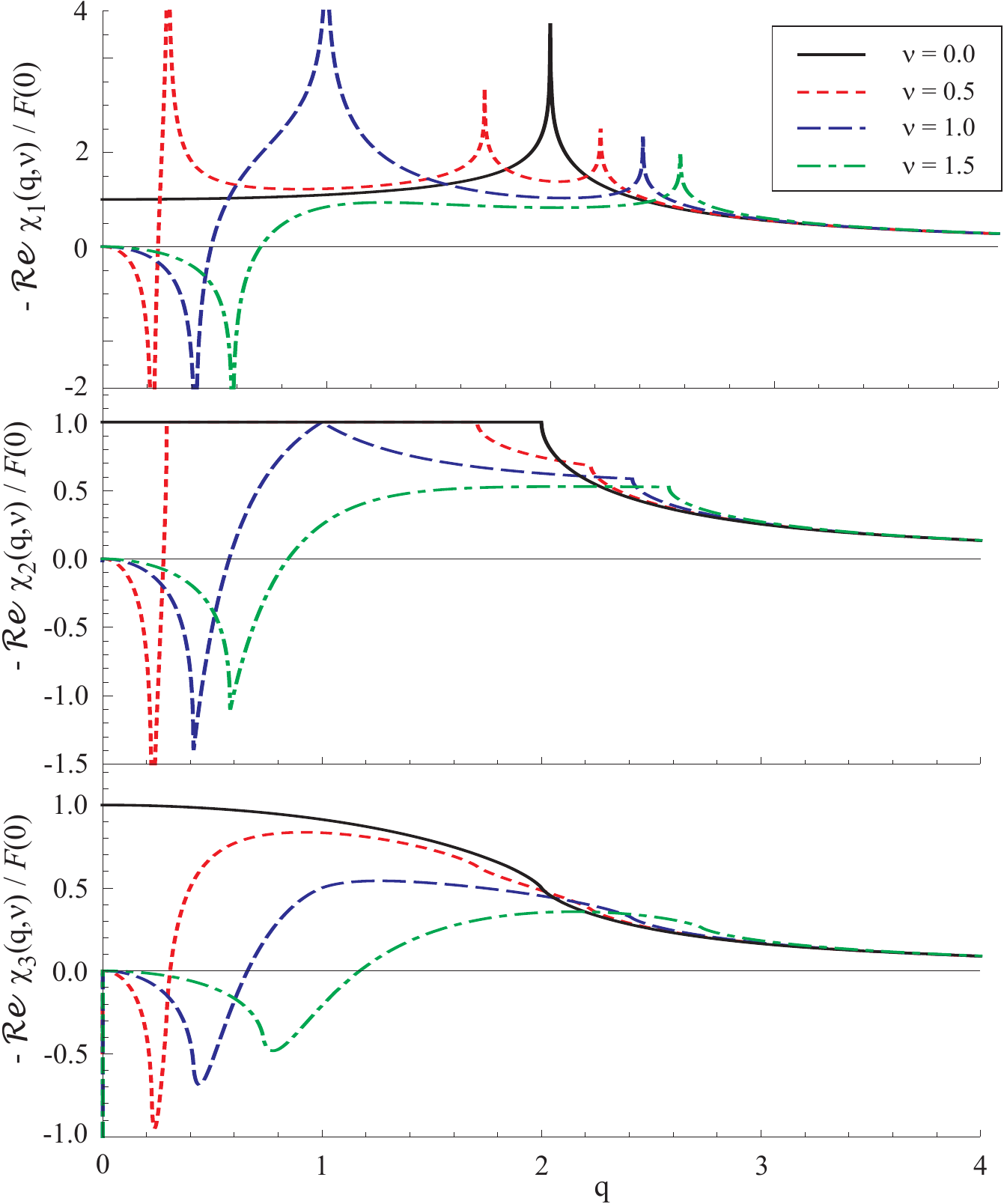}
   \caption{\label{Fig_Re-chi_q} (Color online) Momentum dependence of the real part of the $d$-dimensional Linhard functions, $\mathcal{R}e \, \chi(q,\nu)$, corresponding to several values of the energy transfer, $\nu$. Here, momenta and energies are expressed in rescaled units, i.e. in units of the Fermi momentum, $k_F$, Fermi energy, $\epsilon_F = \hbar^2 k_F^2 / (2m)$.}
\end{figure}
%
%%%%%%%%%%%%%%%%%%%%%%%%%%%%%%%%%%%%%%%%%%%%%%%%%%%%%%%%%%%%%
\begin{figure}[t]
   \centering
   \includegraphics[width=0.9\columnwidth]{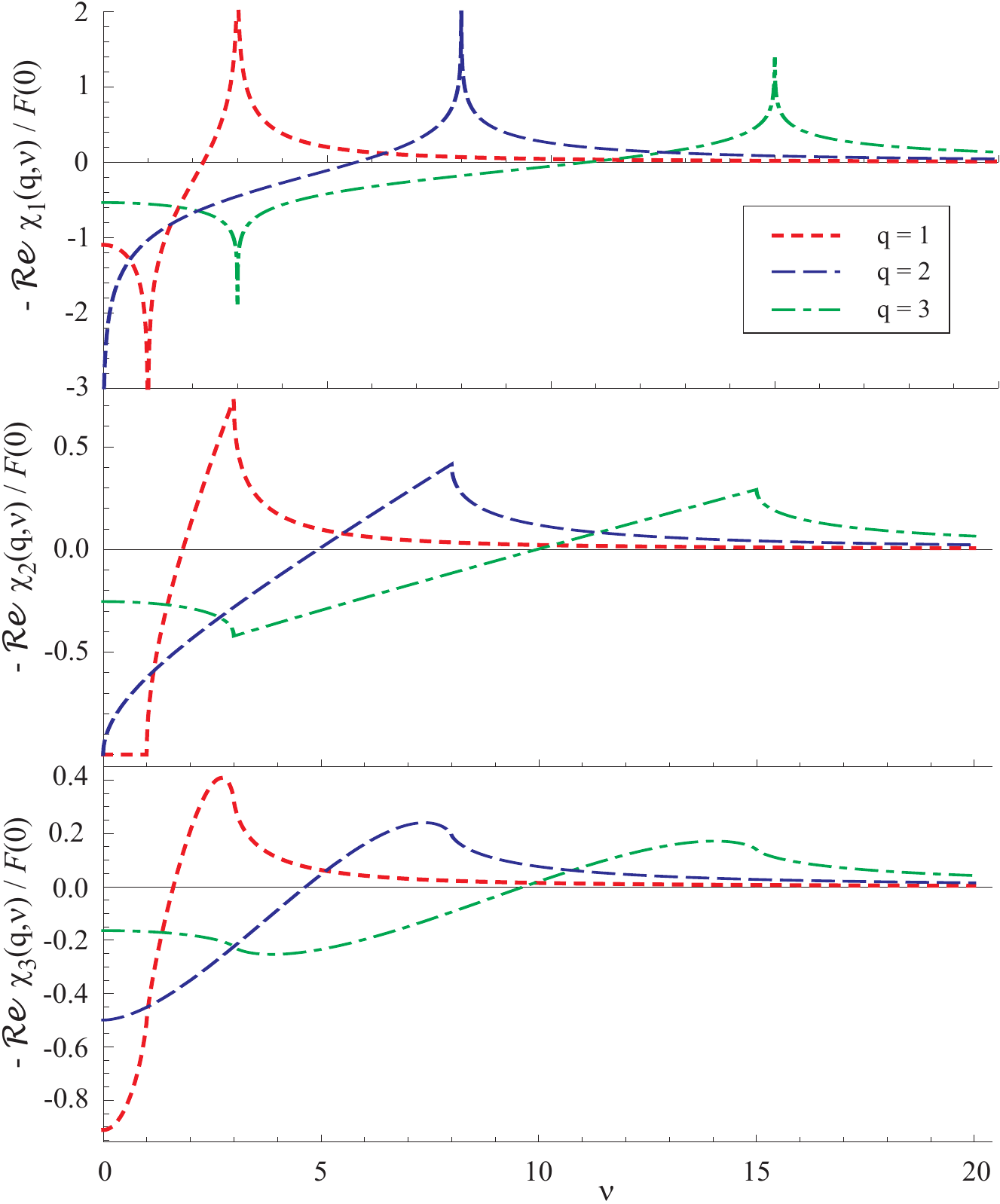}
   \caption{\label{Fig_Re-chi_omega} (Color online) Energy dependence of the real part of the $d$-dimensional Linhard functions, $\mathcal{R}e \, \chi(q,\nu)$, corresponding to several values of the momentum transfer,~$q$. Momenta and energies are expressed in rescaled units.}
\end{figure}
%
%%%%%%%%%%%%%%%%%%%%%%%%%%%%%%%%%%%%%%%%%%%%%%%%%%%%%%%%%%%%%

%
%%%%%%%%%%%%%%%%%%%%%%%%%%%%%%%%%%%%%%%%%%%%%%%%%%%%%%%%%%%%%%%%%%%%%%
\item $d=2$ case:
We have:
\begin{equation}
\mathcal{R}e \, \chi_2^{(\pm)}(q, \nu) =
\frac{k_F^2}{2\pi^2 \epsilon_F} \ \mathcal{P} \! \int_0^1 k \, \mathrm{d}k \,
\int_{0}^{2\pi} \frac{\mathrm{d}\phi}{q_\pm^2 \pm 2 q k \cos\phi}
\>.
\end{equation}
For $|q_\pm^2| / (2q) \ge 1$, the above gives
\begin{align}
& \mathcal{R}e \, \chi_2^{(\pm)}(q, \nu) 
= 
\frac{N_2}{\epsilon_F} \,
\frac{1}{q} \, \mathrm{sgn}( q_\pm^2 )
\int_0^1 \frac{k \, \mathrm{d}k}{\sqrt{ \bigl | q_\pm^2/(2q) \bigr |^2 - k^2}}
\notag \\
& =
\frac{N_2}{\epsilon_F} \ \mathrm{sgn}( q_\pm^2 ) \, 
\biggl [ \, \Bigl | \frac{q_\pm^2}{2q^2} \Bigr | - \frac{1}{q} \, \sqrt{ \bigl | q_\pm^2/(2q) \bigr |^2 - 1} \ \biggr ]
\>,
\end{align}
whereas for $|q_\pm^2| < 2q$, we obtain
\begin{align}
\mathcal{R}e \, \chi_2^{(\pm)}(q, \nu) 
= \, &
\frac{N_2}{\epsilon_F}  \,
\frac{1}{q} \, \mathrm{sgn}( q_\pm^2 )
\int_0^{\frac{1}{2q}q_\pm^2}  \frac{k \, \mathrm{d}k}{\sqrt{ \bigl | q_\pm^2/(2q) \bigr |^2 - k^2}}
\notag \\ = & 
\frac{N_2}{\epsilon_F} \, \frac{q_\pm^2}{2q^2}
\>.
\end{align}
Hence, we obtain the 2D Lindhard function~\cite{stern}:
\begin{align}
\mathcal{R}e & \, \chi_2(q, \nu) 
= 
\frac{N_2}{\epsilon_F} \,  
\Bigl [ 
- 1 
\\ \notag & 
- \mathrm{sgn}(q_-^2) \, \Theta \bigl [ |q_-^2|/(2q) - 1 \bigr ] \ \frac{1}{q} \, \sqrt{ \Bigl | \frac{q_-^2}{2q}  \Bigr |^2 - 1} 
\\ \notag & 
+ \mathrm{sgn}(q_+^2) \, \Theta \bigl [ |q_+^2|/(2q) - 1 \bigr ] \  \frac{1}{q} \, \sqrt{ \Bigl | \frac{q_+^2}{2q}  \Bigr |^2 - 1} 
\ \Biggr ]
\>.
\end{align}

%
%%%%%%%%%%%%%%%%%%%%%%%%%%%%%%%%%%%%%%%%%%%%%%%%%%%%%%%%%%%%%%%%%%%%%%
\item $d=1$ case:
We have:
\begin{align}
\mathcal{R}e \, \chi_1^{(\pm)}(q, \nu)
& 
= \frac{k_F}{\pi \epsilon_F} \ \mathcal{P} \! \int_{-1}^1 
\frac{\mathrm{d}k}{q_\pm^2 \pm 2 q k}
\notag \\ & 
= \frac{N_1}{\epsilon_F} \, \frac{1}{4q} \, 
\log \Bigl | \frac{1 + q_\pm^2/(2 q)}{1 - q_\pm^2/(2 q)} \Bigl |
\>.
\end{align}
The above gives:
\begin{align}
& \mathcal{R}e \, \chi_1(q, \nu) 
\\ \notag & = 
\frac{N_1}{\epsilon_F} \, \frac{1}{4q} \, 
\biggl [ \,
\log \Bigl | \frac{1 + q_-^2/(2 q)}{1 - q_-^2/(2 q)} \Bigl |
\, - \, 
\log \Bigl | \frac{1 + q_+^2/(2 q)}{1 - q_+^2/(2 q)} \Bigl |
\, \biggr ] 
\>.
\end{align}

\end{list}
In Fig.~\ref{Fig_Re-chi_q} we depict the momentum dependence of the real part of the Linhard functions, $\mathcal{R}e \, \chi(q,\nu)$, corresponding to several values of the energy transfer, $\nu$. Similarly, in Fig.~\ref{Fig_Re-chi_omega}  we illustrate the energy dependence of the real part of the Linhard functions, $\mathcal{R}e \, \chi(q,\nu)$, corresponding to several values of the momentum transfer,~$q$. 
%%%%%%%%%%%%%%%%%%%%%%%%%%%%%%%%%%%%%%%%%%%%%%%%%%%%%%%%%%%%%%%%%%%%%%
%

\subsection{Imaginary part of the Lindhard function}

We specialize here to the case of positive energy transfer, $\omega > 0$. Hence, the imaginary part of the Linhard function is given by
\begin{align}
& \mathcal{I}m \, \chi(\qq,\omega)
\\ \notag 
& = - \frac{2\pi}{\hbar} \int_{\le k_F} \frac{\mathrm{d}^dk}{(2\pi)^d} \,
\Theta( |\kk+\qq| - k_F) \, \delta(\omega - \omega_{\kk \qq}) 
\>.
\end{align}
Using rescaled coordinates and after introducing the notation, $\nu=\hbar \omega / \epsilon_F$, we obtain
\begin{align}
& \mathcal{I}m \, \chi(\qq,\nu)
= - \frac{(2\pi) k_F^d}{\epsilon_F} 
\\ \notag 
& \times 
\int_{\le 1} \frac{\mathrm{d}^dk}{(2\pi)^d} \,
\Theta( |\kk+\qq| - 1) \, \delta(\nu - 2 \kk \cdot \qq - q^2) 
\>.
\end{align}
We specialize now to the three-dimensional realizations corresponding to $d=1$, 2 and 3. We have:

%
%%%%%%%%%%%%%%%%%%%%%%%%%%%%%%%%%%%%%%%%%%%%%%%%%%%%%%%%%%%%%%1
\begin{figure}[t]
   \centering
   \includegraphics[width=0.9\columnwidth]{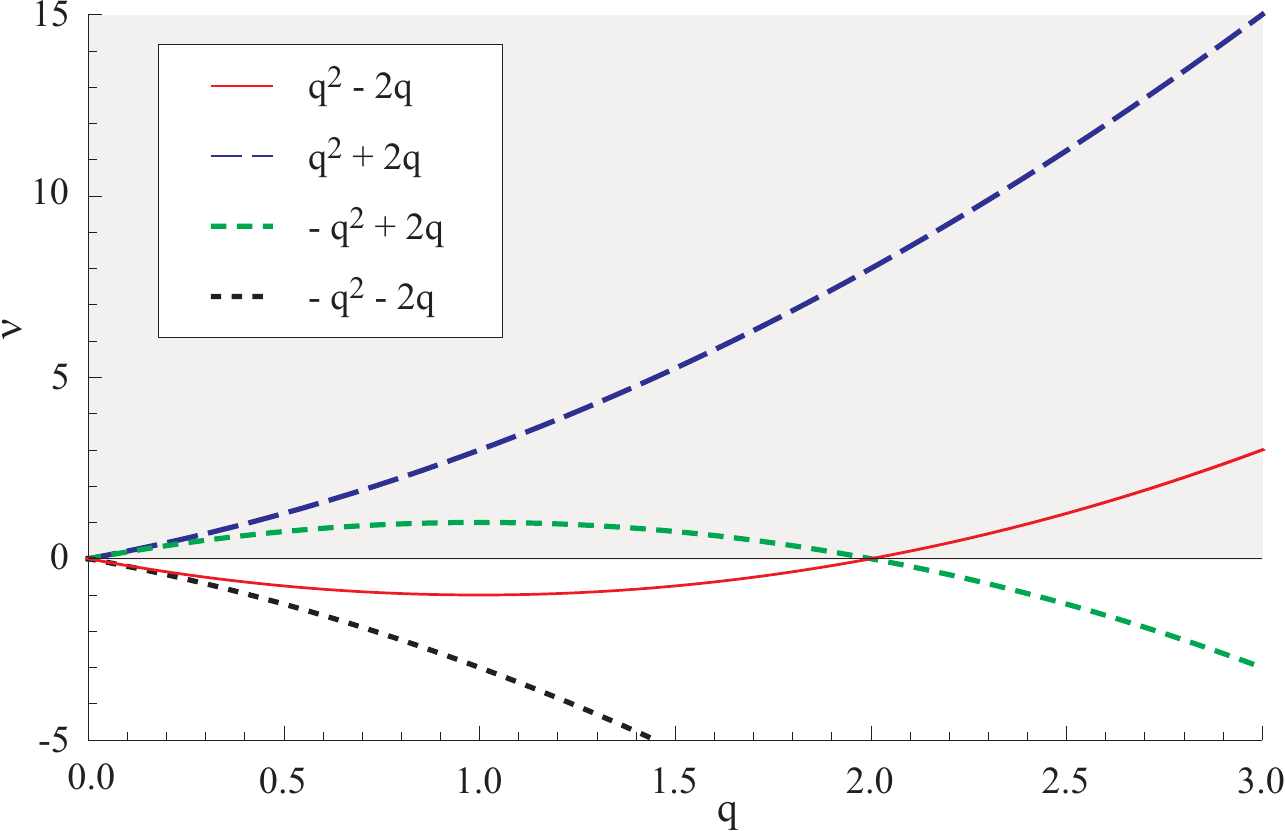}
   \caption{\label{Fig_qs} (Color online) Kinematic constraints in the energy \emph{vs} momentum phase space for the calculation of the imaginary part of the $d$-dimensional Linhard functions. Only positive values of the energy, $\nu$, are physical.  Momenta and energies are expressed in rescaled units.}
\end{figure}
%
%%%%%%%%%%%%%%%%%%%%%%%%%%%%%%%%%%%%%%%%%%%%%%%%%%%%%%%%%%%%%

%
\begin{list}{\labelitemi}{\leftmargin=1em}
%%%%%%%%%%%%%%%%%%%%%%%%%%%%%%%%%%%%%%%%%%%%%%%%%%%%%%%%%%%%%%%%%%%%%%
\item $d=3$ case:
We have:
\begin{align}
\mathcal{I}m \, \chi_3(q,\nu)
= & - \frac{k_F^3}{(2\pi) \epsilon_F} 
\int_0^1 k^2 \, \mathrm{d}k \\ \notag 
& \times 
\int_{-1}^1 \mathrm{d}\mu \
\Theta( |\kk+\qq| - 1) \, \delta(\nu - 2 k q \mu - q^2) 
\>.
\end{align}
We perform the change of variables: $\mu \rightarrow p = |\kk+\qq|$, and obtain 
\begin{align}
& \mathcal{I}m \, \chi_3(q,\nu)
= 
- \, \frac{N_3}{\epsilon_F} \, \frac{3\pi}{2q} 
\\ \notag & \quad \times
\int_0^1 k \, \mathrm{d}k \int_{|k - q|}^{k+q} p \, \mathrm{d}p \
\Theta( p - 1) \, \delta(\nu - p^2 + k^2) 
\>.
\end{align}
Using the identity
\begin{equation}
\delta[f(p)] = \frac{\delta(p-p_0)}{|f'(p_0)|}
\label{eq:delta}
\>, \quad
\mathrm{with}\ f(p_0) = 0 \>,
\end{equation}
we obtain
\begin{align}
& \mathcal{I}m \, \chi_3(q,\nu)
\\ \notag &
= 
- \, \frac{N_3}{\epsilon_F} \, \frac{3\pi}{4q} 
\int_0^1 k \, \mathrm{d}k \int_{|k - q|}^{k+q} \mathrm{d}p \
\Theta( p - 1) \, \delta(p - p_0) 
\>,
\end{align}
where $p_0=\sqrt{\nu + k^2}$. The kinematics conditions require that i) $p_0 \ge 1$, i.e. 
\begin{equation}
k \ge k_a = \sqrt{1-\nu} \>, 
\label{eq:3d_c1}
\end{equation}
and ii) $|k-q|\le p_0 \le k+q$, i.e. 
\begin{equation}
q^2 - 2 k q \le \nu \le q^2 + 2 k q \>,
\label{eq:3d_ene}
\end{equation}
and we recall that $\nu \ge 0$. This gives the energy constrain
\begin{equation}
q^2 - 2 q \le \nu \le q^2 + 2 q
\>.
\end{equation}
The right-hand side of Eq.~\eqref{eq:3d_ene} gives the constraint
\begin{equation}
k \ge k_r =  \frac{1}{2} \Bigl ( \frac{\nu}{q} - q \Bigr ) \>,
\label{eq:3d_c2}
\end{equation}
whereas the left-hand side of Eq.~\eqref{eq:3d_ene} gives
\begin{equation}
k \ge k_l =  -  k_r = - \frac{1}{2} \Bigl ( \frac{\nu}{q} - q \Bigr ) \>.
\end{equation}

%
%%%%%%%%%%%%%%%%%%%%%%%%%%%%%%%%%%%%%%%%%%%%%%%%%%%%%%%%%%%%%
\begin{figure}[t]
   \centering
   \includegraphics[width=0.9\columnwidth]{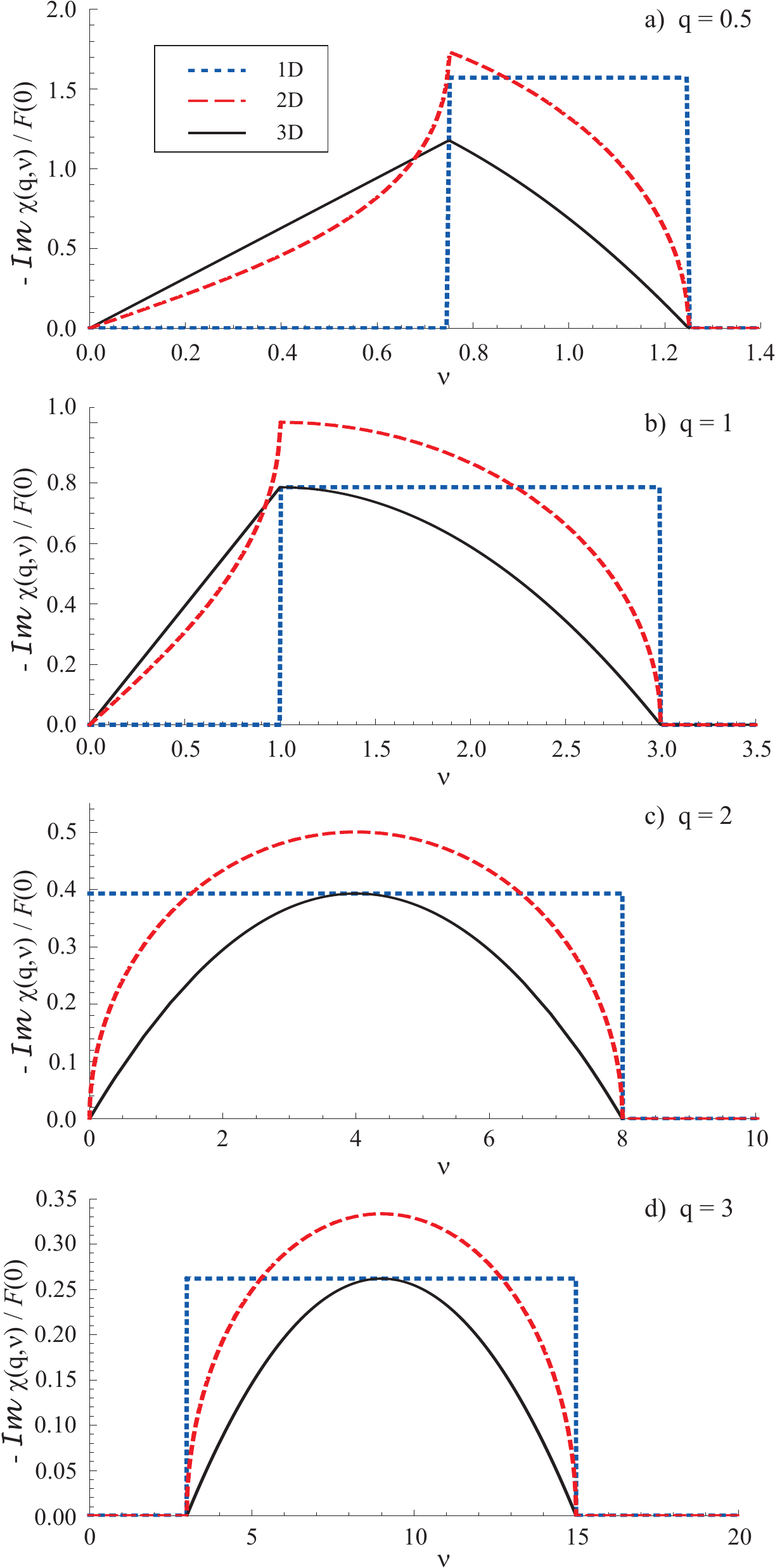}
   \caption{\label{Fig_Im-chi} (Color online) Energy dependence of the imaginary part of the $d$-dimensional Linhard functions, $\mathcal{I}m \, \chi(q,\nu)$, corresponding to several values of the momentum transfer,~$q$.  Momenta and energies are expressed in rescaled units. For convenience, we plot $\frac{1}{2} \, \mathcal{I}m \, \chi_1(q,\nu)$ in the one-dimensional case.}
\end{figure}
%
%%%%%%%%%%%%%%%%%%%%%%%%%%%%%%%%%%%%%%%%%%%%%%%%%%%%%%%%%%%%%

In order to calculate $\mathcal{I}m \, \chi_3(q,\nu)$ we need to find the minimum value of $k$, $k_\mathrm{min} = \max(k_a,k_b)$, where $k_b = \frac{1}{2} \bigl | \frac{\nu}{q} - q \bigr |$. Correspondingly, we introduce the notations
\begin{align}
\mathcal{I}m \, \chi_3^a(q,\nu)
& 
= 
- \frac{N_3}{\epsilon_F} \, \frac{3\pi}{4q} 
\int_{k_a}^1 k \, \mathrm{d}k 
\\ \notag 
& = - \frac{N_3}{\epsilon_F} \, \frac{3\pi}{4} \, \frac{\nu}{2q}
\>,
\end{align}
and
\begin{align}
\mathcal{I}m \, \chi_3^b(q,\nu)
& 
=  - \frac{N_3}{\epsilon_F} \, \frac{3\pi}{4q} 
\int_{k_b}^1 k \, \mathrm{d}k 
\\ \notag &
= - \frac{N_3}{\epsilon_F} \, \frac{3\pi}{4} \, \frac{1}{2q} \,
\Bigl [ 1 - \frac{1}{4} \, \Bigl ( \frac{\nu}{q} - q \Bigr )^2 \Bigr ] 
\>.
\end{align}

We begin by considering the case $k_b \ge k_a$. Using Eqs.~\eqref{eq:3d_c1} and~\eqref{eq:3d_c2}, we obtain the condition
\begin{equation}
\nu^2 + 2 q^2 \nu + q^4 - 4 q^2 \ge 0
\label{eq:f_qnu}
\>,
\end{equation}
which in turn is satisfied if 
\begin{equation}
\nu \ge - q^2 + 2 q
\>.
\label{eq:roots}
\end{equation}
Hence, the relative size of $k_a$ and $k_b$ is obtained by comparing Eqs.~\eqref{eq:3d_ene} and~\eqref{eq:roots}, as illustrated in Fig.~\ref{Fig_qs}.
For $q \ge 2$, the inequality in Eq.~\eqref{eq:f_qnu} is always satisfied because $\nu \ge 0 \ge - q^2 + 2 q$. Therefore, for $q \ge 2$, we have $k_\mathrm{min} = k_b = \frac{1}{2} \bigl |  \frac{\nu}{q} - q \bigr |$. We have
\begin{equation}
\mathcal{I}m \, \chi_3(q \ge 2, q^2 - 2 q \le \nu \le q^2 + 2 q)
= 
\mathcal{I}m \, \chi_3^b(q, \nu)
\>.
\end{equation}

For $q<2$,  the condition $k_b \ge k_a$ is satisfied if the energy satisfies the condition $\nu \ge - q^2 + 2 q$. Then, we obtain
\begin{equation}
\mathcal{I}m \, \chi_3(q<2, - q^2 + 2 q \le \nu \le q^2 + 2 q)
=
\mathcal{I}m \, \chi_3^b(q,\nu)
\>.
\end{equation}
If instead we have $0 \le \nu < - q^2 + 2 q$, then we have, $k_a > k_b$, and the minimum value of $k$ is given as $k_\mathrm{min} = k_a = \sqrt{1-\nu}$. We obtain
\begin{equation}
\mathcal{I}m \, \chi_3(q < 2, 0 \le \nu < - q^2 + 2 q)
=
\mathcal{I}m \, \chi_3^a(q,\nu)
\>.
\end{equation}

%
%%%%%%%%%%%%%%%%%%%%%%%%%%%%%%%%%%%%%%%%%%%%%%%%%%%%%%%%%%%%%%%%%%%%%%
\item $d=2$ case:
We have:
\begin{align}
& \mathcal{I}m \, \chi_2(\qq, \nu)
= - \, \frac{k_F^2}{(2\pi) \epsilon_F} 
\\ \notag 
& \times 
\int_0^1 k \, \mathrm{d}k \int_0^{2\pi} \mathrm{d}\phi \
\Theta( |\kk+\qq| - 1) \, \delta(\nu - 2 k q \cos \phi - q^2) 
\>.
\end{align}
Similarly to the 3D case, we perform the change of variables: $\phi \rightarrow p = |\kk+\qq|$, and obtain 
\begin{align}
& \mathcal{I}m \, \chi_2(q,\nu)
= - \frac{N_2}{\epsilon_F} \, 
\\ \notag & \times \,
4 \!
\int_0^1 k \, \mathrm{d}k \int_{|k - q|}^{k+q} \frac{p \, \mathrm{d}p \,
\Theta( p - 1) \, \delta(\nu - p^2 + k^2)}{\sqrt{\bigl [ (k+q)^2 - p^2 \bigr ] [ p^2 - (k-q)^2 \bigr ] }} 
\>.
\end{align}
Using the identity~\eqref{eq:delta}, we obtain
\begin{align}
\mathcal{I}m \, \chi_2(q,\nu)
= - \frac{N_2}{\epsilon_F} \, 2  
& \int_0^1 \frac{k \, \mathrm{d}k}{\sqrt{4 q^2 k^2 - (\nu-q^2)}} 
\\ \notag & \times 
\int_{|k - q|}^{k+q} \mathrm{d}p \,
\Theta( p - 1) \, \delta(p - p_0)
\>.
\end{align}
where $p_0=\sqrt{\nu + k^2}$. 
The kinematics conditions and the $p$~integral result in a modified range of allowed values for $k$, i.e. $k_\mathrm{min} \le k \le 1$. The discussion regarding kinematics and the value of $k_\mathrm{min}$ is the same as in the 3D case. Hence, we introduce the notations
\begin{align}
\mathcal{I}m \, \chi_2^a(q,\nu)
& 
= - \frac{N_2}{\epsilon_F} \ 2  
\int_{k_a}^1  \frac{k \, \mathrm{d}k}{\sqrt{4 q^2 k^2 - (\nu-q^2)^2}}  
\\ \notag &
= - \frac{N_2}{\epsilon_F} \, \frac{1}{q}
\Biggl [ \, \sqrt{ 1 - \frac{1}{4} \, \Bigl ( \frac{\nu}{q} - q \Bigr )^2} 
\\ \notag & \qquad \qquad \qquad \,
- \sqrt{ \Bigl ( 1 - \frac{q^2}{4} \bigr ) - \frac{\nu}{2} - \frac{\nu^2}{4q^2}} \,
\Biggr ]
\>,
\end{align}
and
\begin{align}
\mathcal{I}m \, \chi_2^b(q,\nu)
& 
= - \frac{N_2}{\epsilon_F} \, 2    
\int_{k_b}^1  \frac{k \, \mathrm{d}k}{\sqrt{4 q^2 k^2 - (\nu-q^2)^2}} 
\\ \notag &
= - \frac{N_2}{\epsilon_F} \, \frac{1}{q} \,
\sqrt{ 1 - \frac{1}{4} \, \Bigl ( \frac{\nu}{q} - q \Bigr )^2} 
\>.
\end{align}
Using the same arguments as in the 3D case, we find 
\begin{equation}
\mathcal{I}m \, \chi_2(q < 2, 0 \le \nu < - q^2 + 2 q)
=
\mathcal{I}m \, \chi_2^a(q,\nu)
\>,
\end{equation}
and
\begin{align}
\mathcal{I}m \, \chi_2(q \ge 2, q^2 - 2 q \le \nu \le q^2 + 2 q)
& =
\\ \notag
\mathcal{I}m \, \chi_2(q<2, - q^2 + 2 q \le \nu \le q^2 + 2 q)
& =
\mathcal{I}m \, \chi_2^b(q,\nu)
\>.
\end{align}

%
%%%%%%%%%%%%%%%%%%%%%%%%%%%%%%%%%%%%%%%%%%%%%%%%%%%%%%%%%%%%%%%%%%%%%%
\item $d=1$ case:
We have:
\begin{equation}
\mathcal{I}m \, \chi_1(q,\nu)
= - \frac{k_F}{\epsilon_F} 
\int_{-1}^1 \!\!  \mathrm{d}k \,
\Theta(k + q - 1) \, \delta(\nu - 2 k q - q^2) 
\>.
\end{equation}
Using the identity~\eqref{eq:delta}, we obtain
\begin{equation}
\mathcal{I}m \, \chi_1(q,\nu)
= 
- \frac{N_1}{\epsilon_F} \, \frac{\pi}{4q} 
\int_{-1}^1 \mathrm{d}k \
\Theta( k_0 + q - 1) \, \delta(k - k_0) 
\>,
\end{equation}
where $k_0= \frac{1}{2} \Bigl ( \frac{\nu}{q} - q \Bigr )$.
The kinematics conditions require that i) $k_0 \ge 1 - q$, i.e. 
\begin{equation}
\nu \ge - \, ( q^2 - 2 q )
\>, 
\label{eq:1d_c1}
\end{equation}
and ii) $-1 \le k_0 \le 1$, i.e. 
\begin{equation}
q^2 - 2 q \le \nu \le q^2 + 2 q \>,
\label{eq:1d_ene}
\end{equation}
and we recall that $\nu \ge 0$. For an energy, $\nu$, satisfying the conditions~\eqref{eq:1d_c1} and~\eqref{eq:1d_ene}, we obtain 
\begin{equation}
\mathcal{I}m \, \chi_1(q,\nu)
= 
- \frac{N_1}{\epsilon_F} \, \frac{\pi}{4q} 
\>,
\end{equation}
We find that the allowed energy domains are $q^2 - 2 q \le \nu \le q^2 + 2 q$ for $q \ge 2$,  and $- \, (q^2 - 2 q) \le \nu \le q^2 + 2 q$ for $0 \le q < 2$.

\end{list}
In Fig.~\ref{Fig_Im-chi} we depict the energy dependence of the imaginary part of the $d$-dimensional Linhard functions, $\mathcal{I}m \, \chi(q,\nu)$, corresponding to several values of the momentum transfer,~$q$. 
%%%%%%%%%%%%%%%%%%%%%%%%%%%%%%%%%%%%%%%%%%%%%%%%%%%%%%%%%%%%%%%%%%%%%%
%
%\vspace{0.2in}

\begin{acknowledgements}
This work was performed in part under the auspices of the United States Department of Energy.  BM would like to acknowledge useful conversations with P.B. Littlewood and D.L. Smith.
\end{acknowledgements}
%%%%%%%%%%%%%%%%%%%%%%%%%%%%%%%%%%%%%%%%%%%%%%%%%%%%%%%%%%%%%%%%%%%%%%

\appendix

\section{Useful integrals}

For the calculation of the FG response function in one and three dimensions we use the integrals:
\begin{align}
\int_{-1}^1 \frac{\mathrm{d}k}{q \pm \alpha k}
= \frac{2}{\alpha} \tanh^{-1} \frac{\alpha}{q}
= \frac{1}{\alpha} \, \log \Bigl | \frac{q + \alpha}{q - \alpha} \Bigr |
\>,
\end{align}
\begin{align}
\int_{-1}^1 \frac{\mathrm{d}k}{q^2 - \alpha^2 k^2}
= \frac{2}{\alpha q} \tanh^{-1} \frac{\alpha}{q}
= \frac{1}{\alpha q} \log \Bigl | \frac{q+\alpha}{q-\alpha} \Bigr |
\>,
\end{align}
\begin{align}
\int_0^1 k \, \mathrm{d}k \, \log | q \pm \alpha k |
= &
- \frac{1}{4}  + \frac{q^2}{2\alpha^2} \log | q |
\\ \notag 
& \pm \frac{q}{2\alpha} 
\pm \frac{\alpha^2 - q^2}{2 \alpha^2} \log | q \pm \alpha |
\>,
\end{align}
and
\begin{align}
\int_0^1 & k \, \mathrm{d}k \,\tanh^{-1} \frac{\alpha k}{q}
= \frac{q}{2\alpha} \Bigl [ 1 + \frac{\alpha^2 - q^2}{2\alpha q} \ \log \Bigl | \frac{q+\alpha}{q-\alpha} \Bigr |
\Bigr ]
\>.
\end{align}

The following integrals are used for the calculation of the FG response function in two dimensions:
\begin{align}
\int_0^{2\pi} \frac{\mathrm{d}\phi}{q - k \cos\phi} 
=  
\biggl \{
\begin{array}{ll}
2\pi \, \mathrm{sgn}(q) \, \bigl ( |q|^2 - k^2 \bigr )^{-\frac{1}{2}} , & |q| > k \>, \\
0, & |q| \le k \>,
\end{array}
\end{align}
\begin{align}
\int_0^{2\pi} \frac{\mathrm{d}\phi}{q^2 - k^2 \cos^2 \phi} 
= 
\Biggl \{
\begin{array}{ll}
\displaystyle{\frac{2\pi}{|q|}} \, \bigl ( |q|^2 - k^2 \bigr )^{-\frac{1}{2}}, & |q| > k \>, \\
0, & |q| \le k \>,
\end{array}
\end{align}
and
\begin{align}
\int_0^1 \frac{k \, \mathrm{d}k}{\sqrt{q^2 - \alpha^2 k^2}}
= \frac{q}{\alpha^2} \, \Bigl [ 1 - \sqrt{1 - (\alpha/q)^2} \Bigr ]
\>,
\end{align}
\begin{align}
\int_0^{\frac{1}{\alpha}q} \frac{k \, \mathrm{d}k}{\sqrt{q^2 - \alpha^2 k^2}}
= \frac{q}{\alpha^2}
\>.
\end{align}

\vfill

%%%%%%%%%%%%%%%%%%%%%%%%%%%%%%%%%%%%%%%%%%%%%%%%%%%%%%%%%%%%%%%%%%%%%%

%%%%%%%%%%%%%%%%%%%%%%%%%%%%%%%%%%%%%%%%%%%%%%%%%%%%%%%%%%%%%%%%%%%%%%

\end{document}